\newcommand{\CLASSINPUTtoptextmargin}{19mm}%
\newcommand{\CLASSINPUTbottomtextmargin}{43mm}%
\newcommand{\CLASSINPUTinnersidemargin}{12.9mm}%
\newcommand{\CLASSINPUToutersidemargin}{12.9mm}%
\documentclass[conference,10pt,a4paper]{IEEEtran}
\usepackage{amsmath}
\usepackage{times}
\usepackage{graphicx}
\usepackage{multirow}
\usepackage[none]{hyphenat}
\usepackage{float}
\usepackage{subfig}
\usepackage{siunitx}
\usepackage[c]{esvect}
\usepackage{hyperref}

\makeatletter

\def\@maketitle{\newpage
\bgroup\par\addvspace{0.5\baselineskip}\centering%
\ifCLASSOPTIONtechnote
   {\bfseries\large\@IEEEcompsoconly{\sffamily}\@title\par}\vskip 1.3em{\lineskip .5em\@IEEEcompsoconly{\sffamily}\@author
   \@IEEEspecialpapernotice\par{\@IEEEcompsoconly{\vskip 1.5em\relax
   \@IEEEtitleabstractindextextbox{\@IEEEtitleabstractindextext}\par
   \hfill\@IEEEcompsocdiamondline\hfill\hbox{}\par}}}\relax
\else
   \vskip0.2em{\EuMWtitlesize\ifCLASSOPTIONtransmag\bfseries\LARGE\fi\@IEEEcompsoconly{\sffamily}\@IEEEcompsocconfonly{\normalfont\normalsize\vskip 2\@IEEEnormalsizeunitybaselineskip
   \bfseries\Large}\@title\par}\vskip1.0em\par
   \ifCLASSOPTIONconference%
      {\@IEEEspecialpapernotice\mbox{}\vskip\@IEEEauthorblockconfadjspace%
       \mbox{}\hfill\begin{@IEEEauthorhalign}\@author\end{@IEEEauthorhalign}\hfill\mbox{}\par}\relax
   \else
      \ifCLASSOPTIONpeerreviewca
         {\@IEEEcompsoconly{\sffamily}\@IEEEspecialpapernotice\mbox{}\vskip\@IEEEauthorblockconfadjspace%
          \mbox{}\hfill\begin{@IEEEauthorhalign}\@author\end{@IEEEauthorhalign}\hfill\mbox{}\par
          {\@IEEEcompsoconly{\vskip 1.5em\relax
           \@IEEEtitleabstractindextextbox{\@IEEEtitleabstractindextext}\par\hfill
           \@IEEEcompsocdiamondline\hfill\hbox{}\par}}}\relax
      \else
         \ifCLASSOPTIONtransmag
           {\@IEEEspecialpapernotice\mbox{}\vskip\@IEEEauthorblockconfadjspace%
            \mbox{}\hfill\begin{@IEEEauthorhalign}\@author\end{@IEEEauthorhalign}\hfill\mbox{}\par
           {\vspace{0.5\baselineskip}\relax\@IEEEtitleabstractindextextbox{\@IEEEtitleabstractindextext}\vspace{-1\baselineskip}\par}}\relax
         \else
           {\lineskip.5em\@IEEEcompsoconly{\sffamily}\sublargesize\@author\@IEEEspecialpapernotice\par
           {\@IEEEcompsoconly{\vskip 1.5em\relax
            \@IEEEtitleabstractindextextbox{\@IEEEtitleabstractindextext}\par\hfill
            \@IEEEcompsocdiamondline\hfill\hbox{}\par}}}\relax
         \fi
      \fi
   \fi
\fi\par\addvspace{0.0\baselineskip}\egroup}

\def\EuMWtitlesize{\@setfontsize{\EuMWtitlesize}{24}{24pt}}
\def\EuMWauthorsize{\@setfontsize{\EuMWauthorsize}{11}{11pt}}
\def\EuMWaffilsize{\@setfontsize{\EuMWaffilsize}{10}{10pt}}
\def\EuMWcaptionsize{\@setfontsize{\EuMWcaptionsize}{9}{10pt}}
\def\EuMWbibsize{\@setfontsize{\EuMWbibsize}{8}{10pt}}

\def\@IEEEauthorblockNstyle{\EuMWauthorsize\@IEEEcompsocnotconfonly{\sffamily}\@IEEEcompsocconfonly{\large}}
\def\@IEEEauthorblockAstyle{\EuMWaffilsize\@IEEEcompsocnotconfonly{\sffamily}\@IEEEcompsocconfonly{\itshape}\@IEEEcompsocconfonly{\large}}
\def\@IEEEauthordefaulttextstyle{\EuMWauthorsize\@IEEEcompsocnotconfonly{\sffamily}\sublargesize}

\def\thebibliography#1{\section*{\refname}%
    \addcontentsline{toc}{section}{\refname}%
    \EuMWbibsize\@IEEEcompsocconfonly{\small}\vskip 0.3\baselineskip plus 0.1\baselineskip minus 0.1\baselineskip
    \list{\@biblabel{\@arabic\c@enumiv}}%
    {\settowidth\labelwidth{\@biblabel{#1}}%
    \leftmargin\labelwidth
    \advance\leftmargin\labelsep\relax
    \itemsep \IEEEbibitemsep\relax
    \usecounter{enumiv}%
    \let\p@enumiv\@empty
    \renewcommand\theenumiv{\@arabic\c@enumiv}}%
    \let\@IEEElatexbibitem\bibitem%
    \def\bibitem{\@IEEEbibitemprefix\@IEEElatexbibitem}%
\def\newblock{\hskip .11em plus .33em minus .07em}%
\ifCLASSOPTIONtechnote\sloppy\clubpenalty4000\widowpenalty4000\interlinepenalty100%
\else\sloppy\clubpenalty4000\widowpenalty4000\interlinepenalty500\fi%
    \sfcode`\.=1000\relax}

\long\def\@makecaption#1#2{%
\ifx\@captype\@IEEEtablestring%
\par\@IEEEtabletopskipstrut
\else
\@IEEEfigurecaptionsepspace
\fi
\setbox\@tempboxa\hbox{\normalfont\footnotesize {#1.}\nobreakspace\nobreakspace #2}%
\ifdim \wd\@tempboxa >\hsize%
\setbox\@tempboxa\hbox{\normalfont\footnotesize {#1.}\nobreakspace\nobreakspace}%
\parbox[t]{\hsize}{\normalfont\footnotesize\noindent\unhbox\@tempboxa#2}%
\else
\ifCLASSOPTIONconference \hbox to\hsize{\normalfont\footnotesize\hfil\box\@tempboxa\hfil}%
\else \hbox to\hsize{\normalfont\footnotesize\box\@tempboxa\hfil}%
\fi\fi
\ifx\@captype\@IEEEtablestring%
\@IEEEtablecaptionsepspace
\else
\fi}

\newlength\tablecaptiontotableskip
\newlength\figuretocaptionskip
\def\@IEEEfigurecaptionsepspace{\vskip\figuretocaptionskip\relax}%
\def\@IEEEtablecaptionsepspace{\vskip\tablecaptiontotableskip\relax}%

\def\abstract{\normalfont%
\@IEEEabskeysecsize\bfseries\textit{\abstractname}\,\bfseries\textit{---}\,%
\@IEEEgobbleleadPARNLSP}%

\def\IEEEkeywords{\normalfont%
\@IEEEabskeysecsize\bfseries\textit{\IEEEkeywordsname}\,\bfseries\textit{---}\,%
\@IEEEgobbleleadPARNLSP}%
\def\endIEEEkeywords{\relax\vspace{0.67ex}%
\par\if@twocolumn\else\endquotation\fi%
\normalsize\normalfont}%

\DeclareRobustCommand*{\EuMWauthorrefmark}[1]{\raisebox{0pt}[0pt][0pt]{\textsuperscript{\footnotesize{#1}}}}%
\def\@IEEEauthorblockNtopspace{0ex}
\def\@IEEEauthorblockAtopspace{1mm}
\def\tablename{Table}
\def\thetable{\arabic{table}}
\def\IEEEkeywordsname{Keywords}
\def\subsubsection{\@startsection{subsubsection}{3}{\z@}{1.5ex plus 1.5ex minus 0.5ex}%
{0.7ex plus .5ex minus 0ex}{\normalfont\normalsize\itshape}}%
\newlength{\CPheadmatchindent}%
\def\@seccntformat#1{\hbox to\CPheadmatchindent{\csname the#1dis\endcsname}\hskip 0.1em \relax}
\IEEEilabelindentA \parindent
\IEEEilabelindent \IEEEilabelindentA
\IEEEelabelindent \parindent
\IEEEdlabelindent \parindent
\IEEElabelindent \parindent
\makeatother

\begin{document}
\raggedbottom
%
%
%
\title{A Realistic Radar Ray Tracing Simulator for Hand Pose Imaging}
%
%
\author{%
\IEEEauthorblockN{%
Johanna Br{\"a}unig\EuMWauthorrefmark{\#1}, 
Christian Sch{\"u}{\ss}ler\EuMWauthorrefmark{\#2}, 
Vanessa Wirth\EuMWauthorrefmark{*},
Marc Stamminger\EuMWauthorrefmark{*},
Ingrid Ullmann\EuMWauthorrefmark{\#}, 
Martin Vossiek\EuMWauthorrefmark{\#3}
}
\IEEEauthorblockA{%
\EuMWauthorrefmark{\#}Institute of Microwaves and Photonics (LHFT), Friedrich-Alexander-Universit{\"a}t Erlangen-N{\"u}rnberg, Germany\\
\EuMWauthorrefmark{*}Chair of Visual Computing (LGDV), Friedrich-Alexander-Universit{\"a}t Erlangen-N{\"u}rnberg, Germany \\
\{\EuMWauthorrefmark{1}johanna.braeunig, \EuMWauthorrefmark{2}christian.schuessler, \EuMWauthorrefmark{3}martin.vossiek\}@fau.de\\
}
}
%
\maketitle
%
%
\begin{abstract}
With the increasing popularity of human-computer interaction applications, there is also growing interest in generating sufficiently large and diverse data sets for automatic radar-based recognition of hand poses and gestures. Radar simulations are a vital approach to generating training data (e.g., for machine learning). Therefore, this work applies a ray tracing method to radar imaging of the hand. The performance of the proposed simulation approach is verified by a comparison of simulation and measurement data based on an imaging radar with a high lateral resolution. In addition, the surface material model incorporated into the ray tracer is highlighted in more detail and parameterized for radar hand imaging. Measurements and simulations show a very high similarity between synthetic and real radar image captures. The presented results demonstrate that it is possible to generate very realistic simulations of radar measurement data even for complex radar hand pose imaging systems.
\end{abstract}
%
%

\section{Introduction}

Automatic gesture recognition using radar-based methods has seen increasing attention as contactless human-computer interaction becomes a more desirable feature in user-friendly interface design \cite{Ahmed.2021}. Despite the increasing interest in static hand poses, little work on this topic has been done in the radar context \cite{Smith.2021}. For example, one attractive area of application is the automated detection of the American Sign Language (ASL) alphabet \cite{CaoDong.2015}. Since machine learning algorithms form the basis of automated detection, a large amount of annotated training data is required. However, recording sufficient training data often presents several problems. First, the measurements and manual annotation of the data are very time-consuming. For this, additional post-processing has to be done, as it is very challenging to generate ground-truth annotations on the same level of accuracy in an automatic manner. Second, it can be difficult to generate a sufficiently diverse data set. Third, the optimal radar hardware may not yet be known or available. With all these limitations, realistic radar response simulations of the human hand are a desirable alternative to real-world captures. 

In \cite{Zhao.2022}, the researchers used the software tool blender and a simple hand model to create synthetic radar responses to hand motions, simulating a radar with low angular resolution. With respect to the radar-based detection of static hand poses, imaging radar systems that incorporate a high lateral resolution, allowing distinction between different fingers, are especially of interest. There are various approaches within computational electromagnetics that can be used to simulate realistic backscattering behavior. Full-wave solvers suffer from extremely high computational burden and long simulation times \cite{Ahmed.2014}. This is why simulation approaches that assume high frequency approximations, such as physical optics (PO) or geometrical optics (GO) are an attractive solution, in case object dimensions are significantly higher than the respective wavelength. In PO, an object surface is first discretized into small patches, followed by the estimation of surface current densities on the basis of surface normals. Then, the radiating field is calculated with an integration over the surface parts. To further reduce computational burden and consider multipath effects, GO which uses rays instead of waves,  can be applied. The concept of shooting and bouncing rays was first introduced in \cite{Ling.1989} to calculate the radar cross section of an open cavity. Here, a dense grid of rays is sent and tracked, taking into account multiple bounces, to calculate the cavity's outgoing field. Subsequently, the scattered field is computed by applying PO and integrating the respective surface currents. However, pure GO-based ray tracing or shooting and bouncing rays approaches can also be used directly to simulate the received baseband signals from imaging radar systems, as shown in \cite{Williams.2015}. Here the authors present a ray tracing approach for whole body imaging systems, assuming that only specular reflections occur at each mesh face. Depending on the wavelength of the radar system and the object's surface roughness, diffuse scattering can be relevant, as it is not efficient to model the roughness by dividing the mesh into a large amount of small surface patches. Therefore, a scattering model based on a bidirectional reflectance distribution function (BRDF) was built on top of a full body radar imaging ray tracing framework in \cite{OrtizJimenez.2017}. However, this BRDF model is rather complex, involving several model parameters that need to be estimated. In \cite{Schasler.2021}, researchers introduced a pure ray tracing approach combined with a simplistic material model to simulate multiple input multiple output (MIMO) radar responses for automotive environments.

In this work, the application of the ray tracing simulation approach from \cite{Schasler.2021} is extended for the high-resolution radar imaging of static hand poses. This involves a more detailed description of the material model's design to simulate the backscattering behavior of the hand. This approach is also evaluated through an appropriate qualitative comparison of simulated hand pose data and measurement of a real human hand. Both simulations and measurements are based on a broadband imaging MIMO radar with 94 transmitting (Tx) and 94 receiving (Rx) antennas, as well as a stepped frequency continuous wave (SFCW) signal form.

\section{MIMO Geometry and Signal Parameters}\label{sec:signal}

\begin{figure}[t!]
	\centering
	\includegraphics[width =0.35\textwidth]{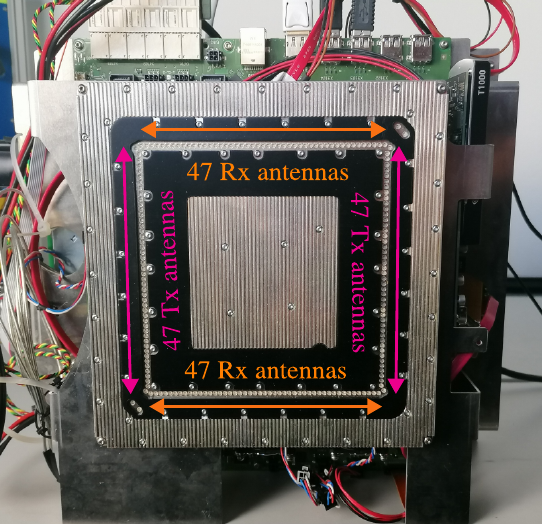}
	\caption{Imaging radar used for the measurement including 94 transmitting (Tx) and receiving antennas (Rx). For a fair comparison, this array geometry was used in all simulations.}
	\label{figure:1}
\end{figure}

The imaging radar used for the measurement in this study, along with its MIMO antenna configuration, is shown in \autoref{figure:1}. The same antenna configuration, consisting of 94 Tx and 94 Rx antennas arranged in a square shape, was used for all simulations. The length of the physical aperture is about \SI{14}{\centi\meter}, resulting in a lateral resolution below \SI{5}{\milli\meter} at a \SI{30}{\centi\meter} distance \cite{Ahmed.2021b}. The distance between adjacent antennas is \SI{3}{\milli\meter}.

In the context of this work, an SFCW waveform, with frequencies ranging from \SI{72}{\giga\hertz} to \SI{82}{\giga\hertz} with $N_\mathrm{f} = 128$ frequency steps, is used for all simulations and measurements.

\section{Simulation Framework}
\label{sec:Framework}

The simulation framework  from \cite{Schasler.2021} consists of a ray tracer that implements a simple surface material model linked to a post-processing system that calculates the radar baseband signals. First, the object geometry is described by a 3D triangle mesh. As a realistic hand model, we incorporated a photogrammetric measurement of one of the author's flat hands. Afterward, the mesh was rigged to allow the synthesis of different hand poses. The generation of hand poses was done using the Autodesk software Maya. The 3D mesh was exported from Maya and imported into the simulation framework, where multiple rays are cast to each mesh triangle and the scattering of each mesh face is simulated.

\subsection{Material Model and Ray Generation}

To model the scattering behavior, a simple yet effective reflection model based on a mixture of two simple material models was applied. One model describes Lambertian or diffuse scattering, and the other simulates specular reflective behavior. The contribution of diffuse scattering was calculated by computing the outgoing ray direction $\vv{t}_\mathrm{o,d}$ as

\begin{equation}
	\vv{t}_\mathrm{o,d} = \vv{n} + \vv{r},
\end{equation}

\noindent where $\vv{n}$ describes the face normal and $\vv{r}$ is a random point on a unit sphere's surface.

To model specularities, the outgoing ray $\vv{t}_\mathrm{o,m}$ was calculated based on the law of reflection,

\begin{equation}
	\vv{t}_\mathrm{o,m} = -2\vv{t_\mathrm{i}}\vv{n},
\end{equation}

\noindent with $\vv{t}_\mathrm{i}$ describing the incident ray. Assuming that $\vv{t}_\mathrm{o,d}$ and $\vv{t}_\mathrm{o,m}$ have unit length, the final outgoing ray was determined by

\begin{equation}
	\vv{t}_\mathrm{o,c} = \alpha \vv{t}_\mathrm{o,d} + (1-\alpha)\vv{t}_\mathrm{o,m}.
\end{equation}

The parameter $\alpha \in [0,1]$ linearly interpolates between a specular ($\alpha = 0$) and a diffuse ($\alpha = 1$) reflection.

Since a diffuse material scatters into various random directions, a single ray reflection per face is not representative enough to gain general results. Hence, in the simple implementation from \cite{Schasler.2021}, each Tx antenna sent out a high number of rays. For detailed descriptions of the ray generation and reception, see \cite{Schasler.2021}.

\subsection{Baseband Signal Simulation}

For each received ray, the total path length $d$ was computed and stored. This information was used to directly simulate the radar baseband signals. As $N_\mathrm{R}$ receptions occur at each Rx antenna, the baseband signal was calculated by coherently summing up the individual signals. This calculation assumed an SFCW signal shape \cite{Iizuka.1984} with $N_\mathrm{f}$ frequency steps. An SFCW baseband signal consists of multiple baseband signals of equal duration, that is, one for each continuous wave (CW) measurement. Thus, for each carrier frequency, the baseband signal $s_\mathrm{b,cw}$ was defined as

\begin{flalign}
	s_\mathrm{b,cw} = \sum_{i=0}^{N_\mathrm{R}} \exp[-\mathrm{j}2\pi(f_0 +& n\Delta f)d_i/c], \\ & n=0,..., N_\mathrm{f} - 1, \nonumber
\end{flalign}

\noindent where $c$ describes the speed of light. The first carrier frequency and the frequency step size are described by $f_0$ and $\Delta f$, respectively.

\section{Radar Response Simulation of Hand Poses}

\begin{figure}[!t]
	\centering
	\subfloat[]{\includegraphics[width =0.4\textwidth]{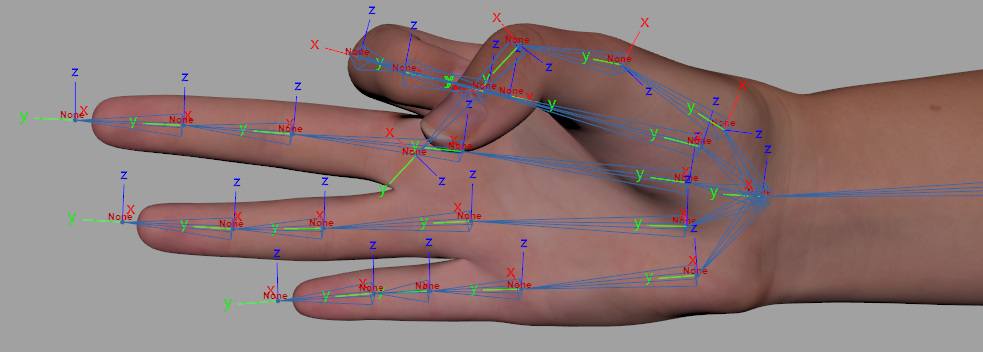}
		\label{figure:2a}}
	\vfill
	\subfloat[]{\includegraphics[width =0.4\textwidth]{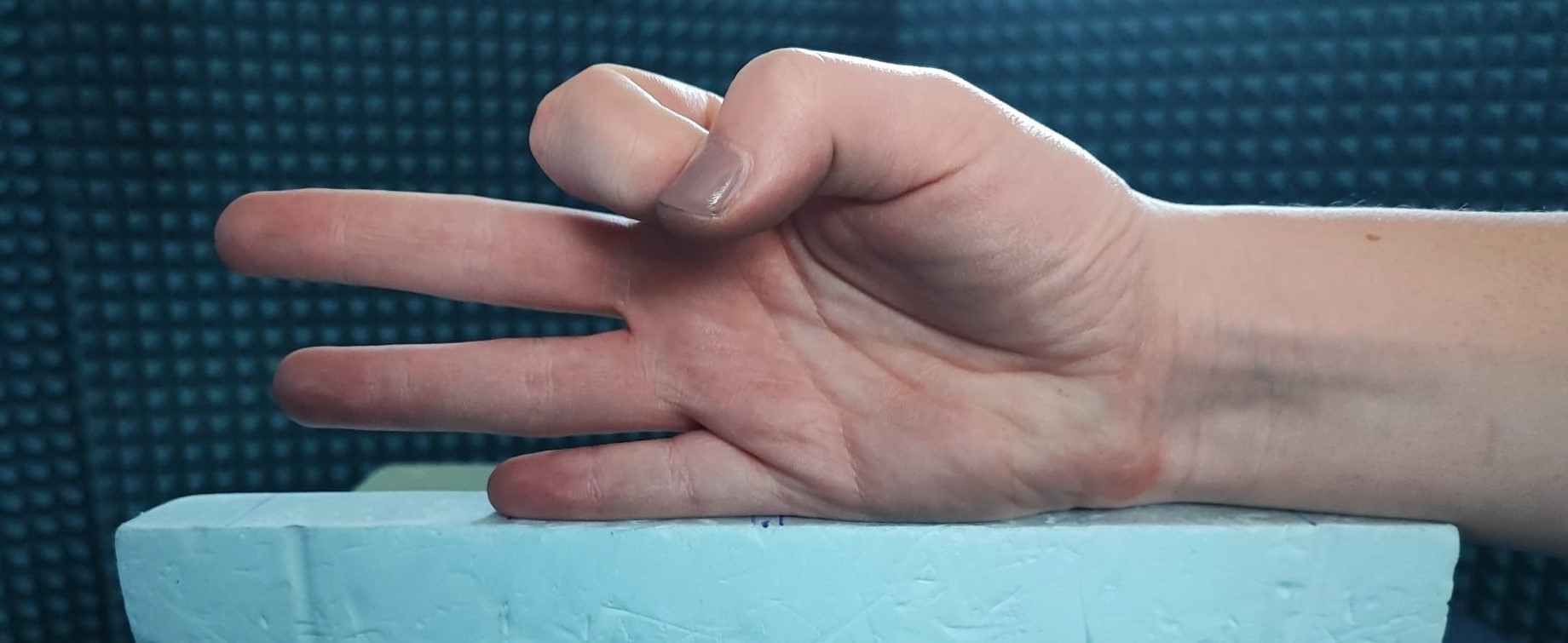}
		\label{figure:2b}}
	\caption{Hand pose for the letter F of American Sign Language alphabet. (a) Synthetic hand pose generation using a rigged mesh. (b) Real hand pose during measurement recordings. The real hand shows skin folds around the knuckles and palm that are not modeled by the 3D mesh.}
	\label{figure:2}
\end{figure}

In \cite{Beckmann.1987b}, the author stated that diffuse scattering components appear when the surface roughness is greater than the wavelength. With a maximum carrier frequency of \SI{82}{\giga\hertz}, the minimum wavelength is approximately \SI{3.7}{\milli\meter}. Irregularities caused by skin folds on the hand are only a few millimeters in size and thus in the range of the minimum wavelength. The reflective behavior is thus mainly characterized by specular reflection. However, the synthesis of skin wrinkles and lifelines with the help of a rigged mesh is challenging, as the hand mesh is limited in its expressiveness. Furthermore, a more expressive hand model results in longer simulation times, as the computation time of the ray tracer is directly proportional to the number of vertices and faces of a mesh. Hence, it is inefficient to synthesize such small details. \autoref{figure:2} illustrates the above-mentioned limitations. It shows the synthetic hand pose used within each of this study's simulations as well as the hand pose imitated by the author's hand during the measurement. In both cases, the letter F of the ASL alphabet was imitated. The rigged mesh did not realistically synthesize the wrinkles and depth variations of the skin surface. However, as skin folds lead to specular reflections in many directions, this work evaluated how skin roughness can be modeled as a partly diffuse scattering process within the simulation tool chain. 

\subsection{Measurement and Simulation Design}\label{Measurement Design}

\begin{figure}[t!]
	\centering
	\includegraphics[width =0.7\columnwidth]{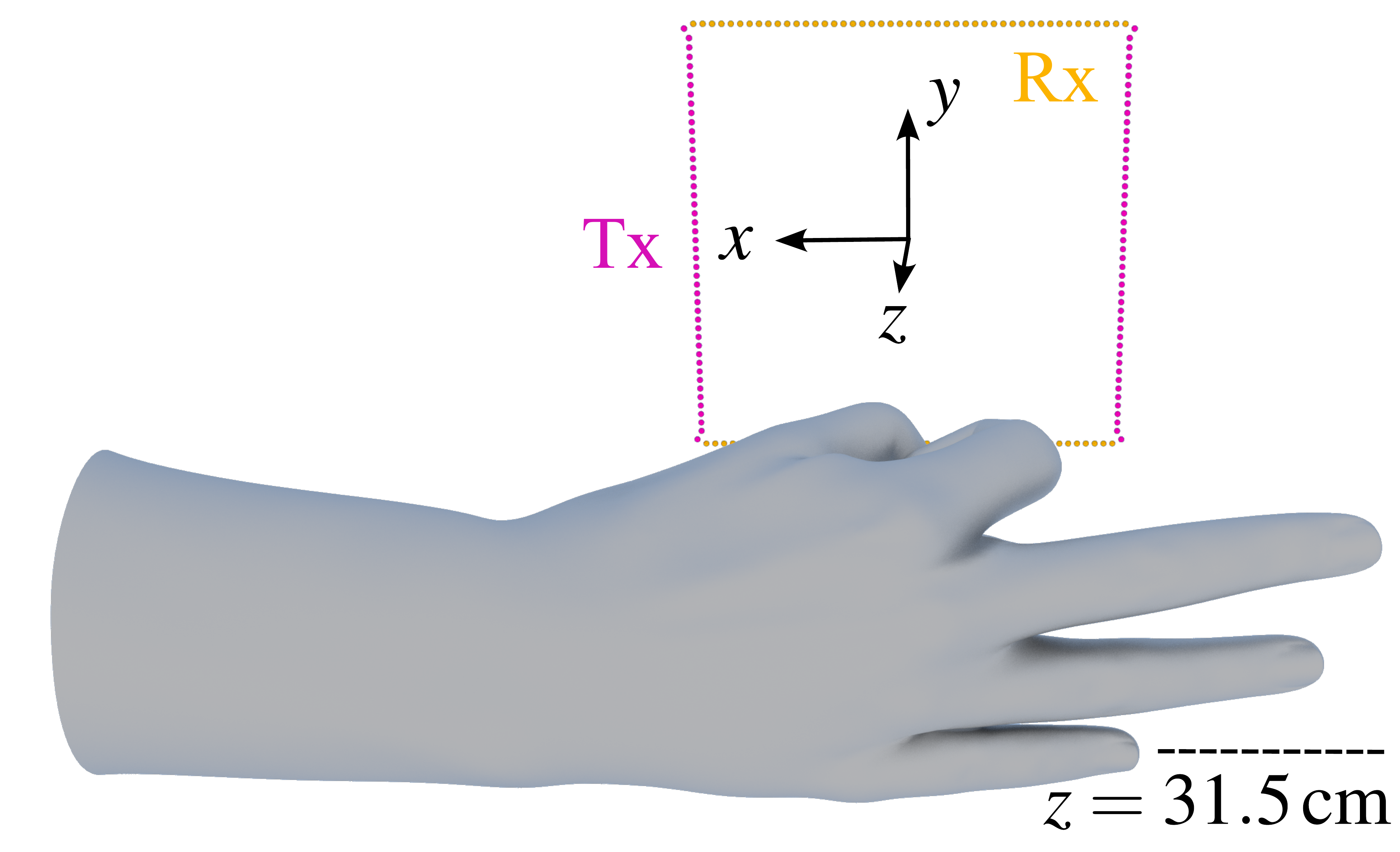}
	\caption{Simulation environment based on the measurement setup.}
	\label{figure:3}
\end{figure}

To allow a comparison between simulation and measurement data, both setups are required to be as similar as possible. For this purpose, as shown in \autoref{figure:2}(b) the hand was positioned on a styrodur block \SI{31.5}{\centi\meter} in front of the MIMO radar visible in \autoref{figure:1}. This allowed the hand to be held as still as possible. Three absorber walls were placed in the back. An SFCW measurement was performed with the signal parameters described in Section \ref{sec:signal}. Based on the recorded radar signal, an SFCW brute-force back-projection algorithm \cite{Ahmed.2021b} was performed to reconstruct the volume of interest, which was in the range of \num{-10} to \SI{10}{\centi\meter} for $x$ and $y$ and \num{26}--\SI{34}{\centi\meter} for the $z$-axis. The voxel dimension of the reconstructed volume was set to \num{1} x \num{1} x \SI{1}{\milli\meter}. A dynamic range of \SI{-15}{\decibel} was applied. By extracting the $z$-coordinate with the maximum amplitude for each pixel in $x-y$, the hand could be located as precisely as possible in 3D space. The 3D information was used to precisely position our hand model within the simulation framework, as depicted in \autoref{figure:3}. Afterward, the simulation was evaluated on different values of $\alpha$, subsequently generating all baseband signals. Finally, the same signal processing as described for the measured data within this section is applied.

\subsection{Comparison of Simulated and Measured Radar Images}

\begin{figure*}[t!]
	\centering
	\includegraphics[width =\textwidth]{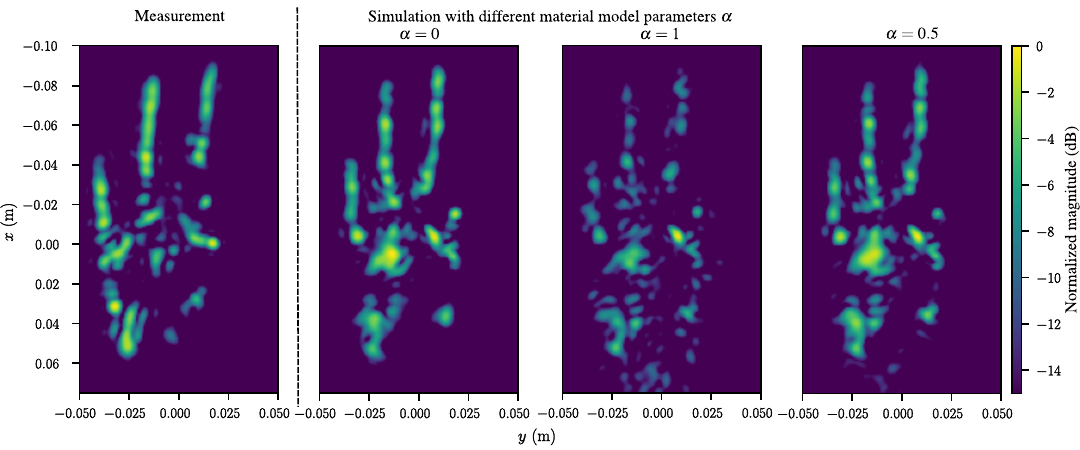}
	\caption{Comparison of measured and simulated radar images of hand pose F for different values of $\alpha$. Specular reflective behavior is simulated by setting $\alpha$ to \num{0}. When $\alpha=1$, the surface parts scatter in a diffuse manner. $\alpha=0.5$ describes a mixed backscattering behavior.}
	\label{figure:4}
\end{figure*}

To compare the measurement and simulation results, each 3D reconstructed volume was reduced to a 2D image by the maximum projection previously described in Section \ref{Measurement Design}. Afterwards, all amplitudes were normalized to the maximum occuring value. The results are depicted in \autoref{figure:4}. In the first two experiments, the value of $\alpha$ was set to the interpolation extents \num{0} and \num{1}, and the respective simulation outcome was compared with the measurement result. The results (see \autoref{figure:4}) show that especially the areas of the extended fingers primarily reflect in a specular manner. Small skin wrinkles in the knuckles do not influence the backscattering behavior. In the palm, the amplitude image is more similar to the result of the diffuse simulation ($\alpha = 1$), as indicated by the small scattering centers. This can be attributed to the deeper skin folds that occur on the real hand due to the hand pose (see \autoref{figure:2}(b)). These are not realistically represented in the simulation model and can be modeled by a partial diffuse backscattering. The last experiment had $\alpha$ set to \num{0.5} to simulate a mixed backscatter behavior. The corresponding amplitude image, which can be seen on the right of \autoref{figure:4}, still yields desirable results for the specular backscattering behavior of the extended fingers and shows several small scattering centers in the palm. The evaluation gradually increased the value of $\alpha$ in increments of \num{0.1}, starting from \num{0}. For an $\alpha$-value equal to or above \num{0.6}, the simulation results for the extended fingers start to significantly deviate from the measurement. Overall, very satisfactory results were obtained in the range of \num{0} to \num{0.5}. 

\section{Conclusion}

In this work, the ray tracing simulation framework from \cite{Schasler.2021} for automotive environments was adapted for radar hand pose imaging. The evaluation was done by comparing suitable simulations with a real measurement based on a 94 Tx/94 Rx MIMO radar. To model reality as accurately as possible, a photogrammetric measurement of one of the author's hands was taken. The comparison shows that the simulation framework is suitable for high-resolution radar hand pose imaging. Furthermore, the design of the associated material model was qualitatively investigated. Within the evaluation, there were multiple valid values of the material model parameter $\alpha$. Realistic simulation results could be generated in the range of \num{0} to \num{0.5}, so a complex BRDF model is not required. The ideal $\alpha$-value also depends on the hand pose and the resulting skin wrinkles. It is also conceivable that additional diversity is created by varying $\alpha$-values when creating  training data sets for machine learning approaches. Furthermore, as a high number of photogrammetric measurements and a subsequent rigging of the meshes is a time-consuming and expensive procedure, the MANO hand model \cite{Romero.2017} can be used instead to generate diverse hand shapes and poses in this regard. In conclusion, the ray tracing simulation framework is very well-suited for the generation oftraining data in radar hand pose imaging. In the future, this framework should be evaluated to simulate the radar responses of realistic hand movements or even whole-body imaging applications.

\bibliographystyle{IEEEtran}
\bibliography{bibliography_EuMW}

\section*{Acknowledgment}

This work was funded by the Deutsche Forschungsgemeinschaft (DFG, German Research Foundation) -- SFB 1483 -- Project-ID 442419336, EmpkinS.
\par The authors would like to thank the Rohde \& Schwarz GmbH \& Co. KG (Munich, Germany) for providing the radar imaging devices and technical support that made this work possible.


\end{document}